\documentclass[10pt,twocolumn]{article}
\usepackage[top=2cm, bottom=2cm, left=1.8cm, right=1.8cm]{geometry}
\usepackage{times}

\usepackage{amsmath,amsthm,amssymb,amsfonts}

\usepackage{ifthen}
\newif\ifshort
\shorttrue   %
\ifshort
	\newcommand{\isShort}{true}
\else
	\newcommand{\isShort}{false}
\fi
\newcommand{\shortVer}[1]{\ifthenelse{\equal{\isShort}{true}}{{#1}}{}}
\newcommand{\longVer}[1]{\ifthenelse{\equal{\isShort}{false}}{{#1}}{}}

\usepackage[compact]{titlesec}
\titlespacing*{\section}{0pt}{*4}{4pt}
\titlespacing{\subsection}{0pt}{*3}{4pt}
\titlespacing{\subsubsection}{0pt}{*1}{1pt}

\usepackage{comment}
\usepackage[keeplastbox]{flushend}
\usepackage{xcolor}
\usepackage{booktabs}
\usepackage{graphicx}
\usepackage{paralist}
\usepackage[small,bf]{caption}
\usepackage{subfig}
\usepackage{multirow}
\usepackage{braket}
\usepackage{mathtools}

\usepackage{color}
\definecolor{darkgreen}{RGB}{47,109,79}
\definecolor{darkblue}{RGB}{57,79,99}
\usepackage{hyperref}
\hypersetup{bookmarks=false, colorlinks=true, plainpages=false, linkcolor=darkgreen, citecolor=darkgreen, urlcolor=darkblue, filecolor=blue}

\renewenvironment{thebibliography}[1]{
  \begin{oldthebibliography}{#1}
    \setlength{\itemsep}{0.0em}
    \setlength{\parskip}{0.0em}
}
{
  \end{oldthebibliography}
}

\usepackage{url}
\usepackage{xspace}

\usepackage{todonotes} %

\newif\ifcomment
\commenttrue
\ifcomment
\newcommand{\XXX}[2]{{\bf \textcolor{yellow}{#1: #2}}}
\newcommand{\nk}[1]{{\bf \textcolor{blue}{NK: #1}}}
\newcommand{\jbnote}[1]{{\bf \textcolor{magenta}{JB: #1}}}
\newcommand{\msnote}[1]{{\bf \textcolor{magenta}{EDC: #1}}}
\newcommand{\il}[1]{{\bf \textcolor{blue}{IL: #1}}}
\newcommand{\gs}[1]{{\bf \textcolor{red}{GS: #1}}}
\newcommand{\enm}[1]{{\bf \textcolor{orange}{EM: #1}}}
\newcommand{\edc}[1]{{\bf \textcolor{blue}{EDC: #1}}}
\newcommand{\gst}[1]{{\bf \textcolor{orange}{GSdT: #1}}}
\else
\newcommand{\XXX}[2]{}
\newcommand{\jbnote}[1]{}
\newcommand{\nk}[1]{}
\newcommand{\msnote}[1]{}
\newcommand{\il}[1]{}
\newcommand{\gs}[1]{}
\newcommand{\edc}[1]{}
\newcommand{\enm}[1]{}
\newcommand{\gst}[1]{}
\fi

\newcommand{\review}[1]{{\textcolor{black}{#1}}}

\newcommand{\descr}[1]{\smallskip\noindent\textbf{#1}}

\newcommand{\dspol}{{\sf{\fontsize{9.5}{10.5}\selectfont /pol/}}\xspace}
\newcommand{\smallpol}{{\sf{\fontsize{8}{9}\selectfont /pol/}}\xspace}

\newcommand{\experimentA}{{{\sc Experiment~1}}\xspace}
\newcommand{\experimentD}{{{\sc Experiment~3}}\xspace}
\newcommand{\experimentE}{{{\sc Experiment~2}}\xspace}

\usepackage{etoolbox}
\makeatletter
\patchcmd\@combinedblfloats{\box\@outputbox}{\unvbox\@outputbox}{}{%
   \errmessage{\noexpand\@combinedblfloats could not be patched}%
}%
 \makeatother

\makeatletter
\def\url@leostyle{%
  \@ifundefined{selectfont}{\def\UrlFont{}}%
  {\def\UrlFont{}}%
}
\makeatother
\usepackage[hyphenbreaks]{breakurl}

\usepackage{etoolbox}

\usepackage[marginal,hang]{footmisc}
\renewcommand{\footnoterule}{%
  \kern -3pt
  \hrule width 1in 
  \kern 2pt
}

\makeatletter
\def\url@leostyle{%
  \@ifundefined{selectfont}{\def\UrlFont{}}%
  {\def\UrlFont{}}%
}
\makeatother  
\makeatletter
\def\url@leostyle{%
  \@ifundefined{selectfont}{\def\UrlFont{}}%
  {\def\UrlFont{}}%
}
\makeatother
\usepackage{xcolor}
\usepackage{flushend}
\usepackage{paralist}
\usepackage{booktabs}
\usepackage{dsfont}

\begin{document}

\sloppy

\title{\bf ``You Know What to Do'': Proactive Detection of YouTube Videos Targeted by Coordinated Hate Attacks\thanks{To appear at the 22nd ACM Conference on Computer-Supported Cooperative Work and Social Computing (CSCW'19).}}
\date{}

\author{
\fontsize{11}{12}\selectfont Enrico Mariconti\\[-0.2ex]
\fontsize{11}{12}\selectfont UCL\\[-0.4ex]
\fontsize{10}{11}\selectfont e.mariconti@ucl.ac.uk
\and \fontsize{11}{12}\selectfont Guillermo Suarez-Tangil\\[-0.2ex]
\fontsize{11}{12}\selectfont King's College London\\[-0.4ex]
\fontsize{10}{11}\selectfont guillermo.suarez-tangil@kcl.ac.uk
\and Jeremy Blackburn\\[-0.2ex]
\fontsize{11}{12}\selectfont Binghamton University\\[-0.4ex]
\fontsize{10}{11}\selectfont blackburn@cs.binghamton.edu
\and Emiliano De Cristofaro\\[-0.2ex]
\fontsize{11}{12}\selectfont UCL \& Alan Turing Institute\\[-0.4ex]
\fontsize{10}{11}\selectfont e.decristofaro@ucl.ac.uk
\and Nicolas Kourtellis\\[-0.2ex]
\fontsize{11}{12}\selectfont Telefonica Research\\[-0.4ex]
\fontsize{10}{11}\selectfont nicolas.kourtellis@telefonica.com
\and Ilias Leontiadis\\[-0.2ex]
\fontsize{11}{12}\selectfont Samsung AI\\[-0.4ex]
\fontsize{10}{11}\selectfont i.leontiadis@samsung.com
\and Jordi Luque Serrano\\[-0.2ex]
\fontsize{11}{12}\selectfont Telefonica Research\\[-0.4ex]
\fontsize{10}{11}\selectfont jordi.luqueserrano@telefonica.com
\and  Gianluca Stringhini\\[-0.2ex]
\fontsize{11}{12}\selectfont Boston University\\[-0.4ex]
\fontsize{10}{11}\selectfont gian@bu.edu}

\maketitle

\maketitle

\begin{abstract}
Video sharing platforms like YouTube are increasingly targeted by aggression and hate attacks. Prior work has shown how these attacks often take place as a result of {\em ``raids,''} i.e., organized efforts by ad-hoc mobs coordinating from third-party communities. Despite the increasing relevance of this phenomenon, however, online services often lack effective countermeasures to mitigate it. Unlike well-studied problems like spam and phishing, coordinated aggressive behavior both targets and is perpetrated by humans, making defense mechanisms that look for automated activity unsuitable. Therefore, the de-facto solution is to reactively rely on user reports and human moderation.

In this paper, we propose an automated solution to identify YouTube videos that are likely to be targeted by coordinated harassers from fringe communities like 4chan. First, we characterize and model YouTube videos along several axes (metadata, audio transcripts, thumbnails) based on a ground truth dataset of videos that were targeted by raids. Then, we use an ensemble of classifiers to determine the likelihood that a video will be raided with very good results (AUC up to 94\%). Overall, our work provides an important first step towards deploying proactive systems to detect and mitigate coordinated hate attacks on platforms like YouTube.
\end{abstract}

\section{Introduction}

Over the years, the Web has shrunk the world, allowing individuals to share viewpoints with many more people than they are able to in real life.
At the same time, however, it has also enabled anti-social and toxic behavior to occur at an unprecedented scale.
As social interactions increasingly take place online, cyber-aggression has unfortunately become a pressing problem~\cite{grigg2010cyber}. %
In particular, coordinated harassment campaigns are more and more frequent, with perpetrators working together to repeatedly target victims with hateful comments~\cite{chatzakou2017hypertext,chess2015conspiracy,elsherief2018peer}.
One example of such behavior is a phenomenon known as {\em raiding}, whereby ad-hoc mobs coordinate on social platforms to organize and orchestrate attacks aimed to disrupt other platforms and undermine users who advocate for issues and policies they do not agree with~\cite{4chan,kumar2018community}.
Abusive activity is generated by humans and not by automated programs, thus, systems to detect unwanted content/bots~\cite{benevenuto2010detecting,nilizadeh2017poised,stringhini2015evilcohort}
are not easily adapted to this problem. 
In fact, Google's CEO, Sundar Pichai, recently identified detecting hate speech as one of the most difficult challenges he is facing~\cite{CNN2019hatespeechyoutube}.
Hence, platforms mostly adopt \emph{reactive} solutions, letting users report abusive accounts and taking actions according to terms of services, e.g., blocking or suspending offenders~\cite{kayes2015ya-abuse}.
However, this approach is inherently slow (as long as seven years for the Sandy Hook massacre videos~\cite{CNN2019hatespeechyoutube}), and limited by biases in the reports and by the resources available to verify them. %
In this paper, we focus on {\em raids against YouTube videos}.
We do so for the following reasons:
(1)~YouTube is one of the top visited sites worldwide, with more than 1 billion users and 1 billion hours of videos watched every day\footnote{\url{https://www.youtube.com/yt/about/press/}}, and (2)~it is plagued by aggressive behavior and extremism~\cite{newrepublic}.
In fact, previous work~\cite{4chan} has shown that YouTube is the most heavily targeted platform by hateful and alt-right communities\longVer{within 4chan}, and in particular 4chan's Politically Incorrect board (\dspol).
Besides providing a characterization that identifies when a raid has occurred, however, previous work has not provided solutions to mitigate the problem.

In this paper, we propose a \emph{proactive} approach towards curbing coordinated hate attacks against YouTube users. 
Rather than looking at attacks as they happen, or at known abusive accounts, {\bf\em we investigate whether we can automatically identify YouTube videos that are likely to be raided.} 
We present a system that relies on multiple features of YouTube videos, such as title, category, thumbnail preview, as well as audio transcripts, to build a model of the characteristics of videos that are commonly raided.
This also allows us to gain an understanding of {\em what} content attracts raids, i.e., \emph{why} these videos are raided.

Our experiments rely on ground truth dataset of 428 raided YouTube videos obtained from our previous work~\cite{4chan}, comparing them to 15K regular YouTube videos that were not targeted by raiders.
Based on our analysis, we build classification models to assess, at upload time, whether a video is likely to be raided in the future. 
We rely on an {\em ensemble} of classifiers, each looking at a different element of the video (metadata, thumbnails, and audio transcripts), 
and build an ensemble detection algorithm that performs quite well, reaching AUC values of up to 94\%. 
Overall, our work provides an important first step towards curbing raids on video sharing platforms, as we show how to detect videos targeted by coordinated hate attacks. 

In summary, our paper makes the following contributions: %
\begin{compactenum}
\item We analyze and model YouTube raids perpetrated by users of 4chan, using a ground truth dataset of 428 raided videos. %
\item We build an ensemble classification model %
  geared to determine the likelihood that a YouTube video will be raided in the future, using a number of features (video metadata, audio transcripts, thumbnails). Our system achieves an AUC of 94\% when analyzing raided videos posted on 4chan with respect to all other non raided videos in our dataset.
\item We provide concrete suggestions as to how video platforms can deploy our methodology to detect raids and mitigate their impact. %
\end{compactenum}

\section{Background \& Related Work} \label{sec:preliminaries}

Hate attacks on online services can happen in a number of ways.
In this paper, we focus on organized attacks -- {\em ``raids''} -- which are orchestrated by a community 
and target users on other platforms~\cite{4chan,kumar2018community}.
In this section, we provide an overview of online raids, and describe how fringe communities organize and orchestrate them.
Then, we review relevant prior work.

\subsection{Anatomy of Online Raids} \label{sec:raids}

Unlike ``typical'' attacks on online services, such as denial of service~\cite{rossow2014amplification}, a raid is an attack on the community that calls a service home.
The goal is not to disrupt the service itself, but rather to cause chaos and disruption to the users of the service.
As such, online raids are a growing socio-technical problem.
Nonetheless, it is hard to provide a precise definition of them.
In the following, we offer a description of them based on previous work as well as our own observations of raids in the wild.

A prototypical raid %
begins with a user finding a YouTube video and posting a link to it on a third party community, e.g., 4chan's /pol/.
In some cases, the original poster, or another user, might also write comments like {\em ``you know what to do.''}
Shortly after, the YouTube video starts receiving a large number of negative and hateful comments.
Overall, raids present a couple of key characteristics.
For instance, they typically attract a large number of users, joining an effort to explicitly disrupt any productive/civil discourse that might be occurring.
This is different from what would normally happen with possibly controversial videos (say, e.g., a video posted by social justice advocates), which also attract opposing points of view, though organically.
The raid often generates a sudden, unexpected attack by uninvolved users.

\longVer{
\begin{figure}[t]
\centering
\includegraphics[width=0.8\columnwidth]{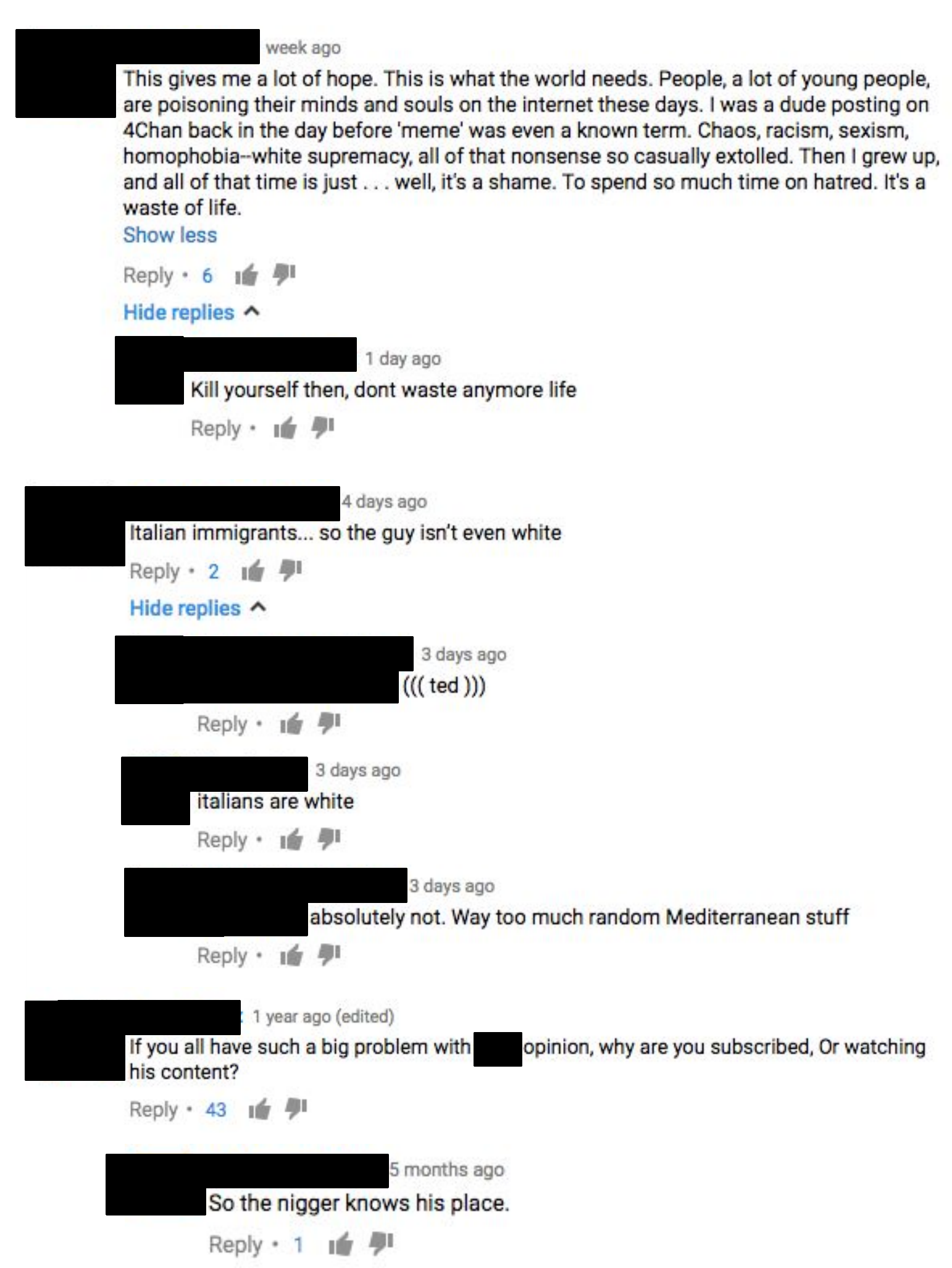}
\caption{Example of comments from raided YouTube videos, with usernames and profile pictures removed for the sake of privacy.}
\label{fig:raid_example}
\end{figure}
}

\longVer{
Consider, for example, the comments from a few raided videos showed in Figure~\ref{fig:raid_example}.
The first comment specifically calls out the racist ideology espoused on 4chan (and based on the comparative analysis between different 4chan boards from~\cite{4chan}, this commenter is likely referring to \dspol in particular).
Another user---presumably a 4chan user---responds by telling the commenter to kill themselves.
The next set of comments refers to the racist meme that Italians are not ``white.''
The first response also  uses the anti-semitic triple parenthesis meme\footnote{\url{https://en.wikipedia.org/wiki/Triple_parentheses}}%
 to imply that the TED organization (the source of the video being commented on) is  a tool of a supposed Jewish conspiracy.
When another user responds that Italians are in fact white, the original commenter provides a justification for his assertion: that Italians are too Mediterranean to be considered white.
In the final set of comments, a user asks why the ``raiders'' are even watching the video if they have issues with the poster's opinions;
the response is that they need to ensure that the video's poster (a minority) ``knows his place.''}

Another characteristic of raids is their semi-coordinated nature.
While a sudden increase in hateful comments to a video is obvious to an outside observer, what is not obvious is the fact that these comments are part of a coordinated attack.
In fact, those participating in a raid may even discuss the ``fruits of their labor'' on the third party site that they organize on.
For example, as discussed in~\cite{4chan}, \dspol threads serve as an aggregation point for raiders;
users will post a hateful comment to the targeted YouTube video, and then brag about it on \dspol.
This observation led the authors to identify videos that might have been targeted by a raid by measuring the number of ``hate comments per second'' (HCPS),
as well as the synchronization between the comments posted on the YouTube video and those appearing on the \dspol thread advocating for a raid.

\begin{figure}[t]
 \centering
 \includegraphics[width=0.75\columnwidth]{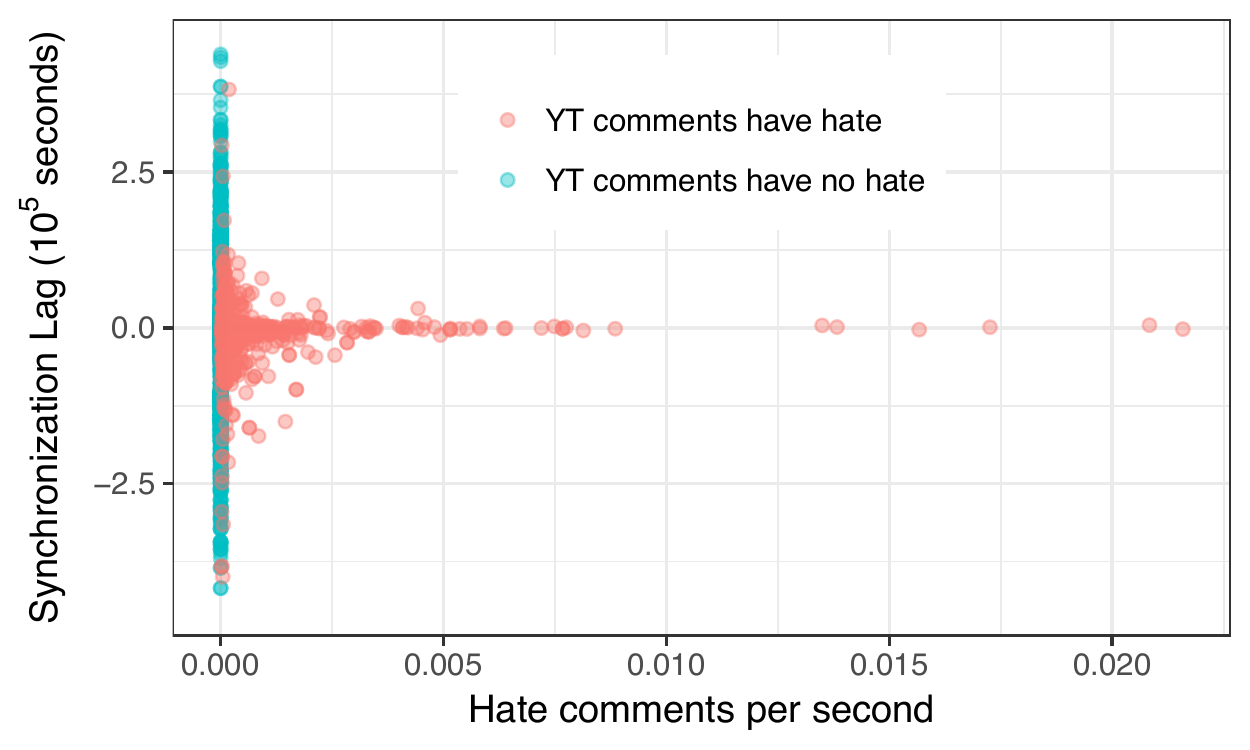}
  \vspace{-0.2cm}
  \caption{Distribution of videos from~\cite{4chan},  according to the synchronization of their comments with the 4chan thread where the URL was posted and the number of hate words that appear in the comments.}
  \label{fig:hcps} 
\end{figure}

By correlating the synchronization lag and the HCPS metric, Hine et al.~\cite{4chan} 
identified a set of videos targeted by raids during their observation period.
This approach was validated by showing an increase in the overlap of YouTube accounts between videos as the synchronization lag decreases: the same accounts were more likely to be involved in the YouTube comments.
In other words, it was not random YouTube users leaving hateful comments, but rather ``serial'' raiders almost assuredly originating from \dspol.
In Figure~\ref{fig:hcps}, we show the distribution of videos in dataset from~\cite{4chan}
according to synchronization lag and HCPS: observe that, the closer the lag is to zero, the higher rate of hate comments received by the video.

\subsection{Related Work}\label{sec:related-work}

\descr{YouTube.} YouTube is used every day by millions of individuals to share various kinds of videos, e.g., music, lectures, gaming, video blogs, etc.~\cite{park2014youtube}.
Organized groups are also active on the platform; some use it to reach and grow their network of support and organize activities, while others to propel radicalization and initiatives against other groups.
Previous work has looked at the use of YouTube by LGBT users for self-disclosure~\cite{green2015lgbt}, for anti- or pro-anorexia \cite{oksanen2015anorexia}, fat stigmatization \cite{hussin2011fat-stigma}, sharing violent content \cite{weaver2012violentvideo}, far-right propaganda \cite{ekman2014right-wing}, as well as  Jihad and self-radicalization \cite{conway2008jihad}.
These topics often attract considerable attention and typically lead to unwanted behavior studied, e.g., in the context of swearing, abusive, hateful, or flaming comments, as well as activity against the video poster and its supporters \cite{moor2010flaming,jonson2013flaming}.

\descr{Hate on Social Platforms.} \review{Hate has driven many historical moments in the past. Even by focusing only on what happened online, researchers have studied hateful content already in the early 2000s, analyzing chat rooms behavior~\cite{glaser2002studying}, extremist websites~\cite{gerstenfeld2003hate}, and blogs~\cite{chau2007mining}. As social networks became global communities, they also amplified discussions, creating more controversy and conflict~\cite{sobkowicz2010dynamics}.}
\review{On social networks, hate has been studied mainly by following politics and, in particular, the analysis of alt-right communities~\cite{ben2016hate,zannettou2018gab} and populism~\cite{gerbaudo2018social}.} %

Interactions in social communities~\cite{zhang2018characterizing} and between different groups~\cite{datta2017identifying} often leads to conflicts. 
In some cases, this may involve trolling~\cite{cheng2017anyone,israni2017snitches} or harassment~\cite{vitak2017identifying}. 
Researchers have overall studied bullying and hate speech on, e.g., Twitter \cite{burnap2016twitter-hate,chatzakou2017hypertext} or Yahoo News and Finance %
\cite{nobata2016abusivelanguageYahoo}.
Also, Salminen et al.~\cite{salminen2018anatomy} focus on a single target across platforms while Olteanu et al.~\cite{olteanu2018effect} look at the hateful speech on Twitter and Reddit in relation to extremist violence. 
Mostly, social networks have reacted to this behavior through bans (e.g., on Reddit~\cite{chandrasekharan2017you}) and blacklisting users (e.g., on Twitter~\cite{jhaver2018online}). Whereas, we aim to build a proactive, rather than reactive, system.
Overall, hate speech detection systems have become a prominent area of research in the last years~\cite{elsherief2018hate,Soni:2018:evil,Maity:2018:Incivility}.
By contrast, we focus on hateful activity against YouTube videos, and in particular on studying the types of videos that are more likely to be targeted by attacks.

\descr{Cyberbullying \& Aggression on YouTube.} Prior work has also studied controversial YouTube videos aiming to understand what types of comments and user reaction certain categories of videos attract.
For instance, Alhabash et al.~\cite{alhabash2015civicbehavioralintentions} measure civic behavioral intention upon exposure to highly or lowly arousing videos showing cyberbullying activity, while Lange~\cite{lange2014ranting} studies user-posted ``ranting'' videos, which, although appearing to be aggressive, actually cause subtle differences in how they engage other users.
Recent studies also analyze YouTube videos' comments to detect hate speech, bullying, and aggression via swearing in political videos.
Kwon et al.~\cite{kwon2017trumpaggresion,kwon2017aggressioncontagion} investigate whether aggressive behavior (in online comments) can be contagious, observing mimicry of verbal aggression via swearing comments against Donald Trump's campaign channel.
Interestingly, this aggressive emotional state can lead to contagious effects through textual mimicry.
In this paper, we build on previous work on characterizing raiding behavior on YouTube, presenting a data-driven approach to identify \emph{which} videos are likely to be the target of a raid.

\descr{Detection.} Another line of work has looked at offensive/harmful YouTube videos and how to automatically detect them.
This is an orthogonal problem to ours, as we look at videos posted with a legitimate purpose, that are later victim of coordinated attacks.
Sureka et al.~\cite{sureka2010extremistyoutube} use social network analysis %
to identify extremist videos on YouTube, while Aggarwal et al.~\cite{aggarwal2014youtubeharassment} detects violent and abusive videos, by mining the video's metadata such as linguistic features in the title and description, popularity of video, duration and category.
Finally, Agarwal and Sureka~\cite{agarwal2014youtubecrawler} search for malicious and hateful videos using a topical crawler, best-first search, and shark-search for navigating nodes and links on YouTube.

Marathe and Shirsat~\cite{marathe2015youtubebullying} study detection techniques used for other problems, e.g., spam detection, and assess whether they could be applied to bullying detection.
Also, Dadvar et al.~\cite{dadvar2014youtubebullying} use machine learning to detect YouTube users exhibiting cyberbullying behavior.
Whereas, rather than focusing on single offending users, we look at videos that are likely to receive hate and raids and their attributes from various users.
Finally, Hine et al.~\cite{4chan}, as already discussed, show that underground forums such as 4chan %
organize raids to platforms like Twitter, Google, and YouTube.

\descr{Remarks.} In conclusion, to the best of our knowledge, this work is the first to study video properties, such as their transcript, metadata, and thumbnail, to shed light on the characteristics of the videos raided by the users of such platforms, using advanced machine and deep learning techniques to perform detection of videos targeted by raids.

\section{Datasets} \label{preliminaries:data}
In this section, we introduce our three datasets used throughout the paper, as also summarized in Table~\ref{tab:datasets}:
\begin{compactenum}
\item Videos raided after being posted on \dspol, as identified by~\cite{4chan};
\item Videos posted on \dspol which were {\em not} raided;
\item Random YouTube videos, which we use to compare raided videos against.
\end{compactenum}

\begin{table}[t]
\small
\centering
\begin{tabular}{llr}
\toprule
{\bf Type} & {\bf Source} & {\bf \# Videos} \\
\midrule
Raided 		& 4chan (\smallpol)	&	428 \\ 
Non-Raided 	& 4chan (\smallpol)	&  	789 \\ 
Random 		& YouTube			& 	14,444 \\ 
\bottomrule
\end{tabular}
\vspace{-0.2cm}
\caption{Overview of our datasets of YouTube videos. ``Source'' denotes the platform from where the link to the YouTube video was collected.}\label{tab:datasets}
\end{table}

\descr{Raided videos posted on 4chan (ground truth).} %
We start by collecting a ground truth dataset of raided YouTube videos.
As discussed previously, fringe communities within 4chan are often responsible for organizing raids against YouTube users that promote ideas that they do not agree with.
Raiders attack such videos, and being part of a group make them feel authorized to express their point of view by insulting the other users and disrupting the civil discussion on the topic.
Therefore, we obtain the dataset of YouTube links posted on 4chan over a 2.5-month period in 2016 (June to mid September) from the authors of~\cite{4chan}.
For our purposes, we want to choose \emph{conservative} thresholds to ensure we only select videos that we are confident were raided.
Thus, based on Figure~\ref{fig:hcps}, we select videos with $HCPS>10^{-4}$ and time lag less than a day, resulting in 428 videos (out of 1,217) that were raided. [See Section~\ref{sec:raids} for details on the Hate Comments Per Second (HPCS) metric.]
We manually examined this ground truth dataset to further increase our confidence that they were indeed raided.

\descr{Non-raided videos posted on 4chan.} %
Although many YouTube videos posted on 4chan's \dspol are raided, obviously not all videos posted attract hateful behavior.
Figure~\ref{fig:hcps} shows that videos that have a high lag compared to the thread in which they are posted are unlikely to see much hateful behavior.
To compare the characteristics of these videos to the raided ones, we build a second dataset with videos that were posted on 4chan but were \emph{not} raided.
We use conservative thresholds to ensure that we do not mistakenly included raided videos: to be part of this set, a video needs to have both a synchronization lag of more than one day compared to the 4chan thread it was posted in, and to have a HCPS of 0. These choices leave an unselected set of videos from the \dspol database, making our sets cleaner and more accurate.
This yields 789 non-raided videos.
\descr{Random YouTube videos.} %
Finally, in order to draw comparisons with the ground truth of raided videos, we need to collect a set of YouTube videos that are likely not raided \review{(we may refer to them as {\em not raided} from now on)}.
We use the YouTube API and download 50 of the top videos across a variety of categories.
In the end, we collected 14,444 videos, selected following the same distribution of (YouTube) categories as those linked on \dspol. We downloaded videos during the same time window as the ones posted on \dspol to build a random set as reliable as possible, however, the dynamics of the raids may force our system to periodically update the training set.

\descr{Ethical considerations.} Our research protocol received ethics approval from University College London.
We also followed standard ethical practices to minimize information disclosure for all datasets, e.g., we discarded {\em any} personal information about the users uploading or commenting on the videos, encrypted data at rest, etc.

\section{Video Processing and Analysis} \label{sec:characteristics}

We now present the methods used to analyze the characteristics of the YouTube videos in our dataset that received raids.
We look at the metadata of a video, its audio transcript, as well as the thumbnail preview.
We then use the insights derived from this analysis to build a classifier geared to determine whether a YouTube video is likely to receive a raid.  

\subsection{Metadata}\label{preliminaries:metadata}
In addition to the actual videos, we also collect the associated metadata, specifically: title, duration, category, description, and tags.
Except for the duration, these fields are entered by the user uploading the video. %
Naturally, title, duration, and description often play a major role in a user's decision to watch the video as they are the first elements that they see.
Also, the tags provide an intuition of a video's topics, and are actually also used by YouTube to suggest other videos---in a way, watching a suggested video might actually trigger a post on 4chan.
Looking at the category for videos posted on 4chan, we anecdotally find that many of them include news, politics, and ethnic issues.
\descr{Evidence of controversial topics.}
We perform term frequency-inverse document frequency (\emph{TF-IDF}) analysis on the string metadata (title, tags, and description) to extract information about the most used keywords in the different groups of videos, finding that random videos often include ``Google,'' ``music,'' and ``love'' (top 3 used words), as well as ``follow'' and ``subscribe.''
By contrast, all videos posted on 4chan include politics-related words such as ``Hillary'' and ``Trump,'' or indications of racial content like ``black,'' while only raided videos have certain words like ``police,'' ``lives'' (likely related to the Black Lives Matter movement), or ``Alex'' (referring to InfoWars' Alex Jones, a conspiracy theorist, who is well known in alt-right circles).\footnote{\url{https://en.wikipedia.org/wiki/Alex_Jones}}

The differences in the topics used in the metadata are extremely important: search engines are affected by the content of the metadata, especially tags; moreover YouTube suggests videos to the user based on many of these fields. 
Overall, we observe that random YouTube videos have few topics in common with the 4chan videos, while there are some similarities between the set of videos posted on 4chan but not raided and those that were raided.%
\subsection{Audio Processing}\label{preliminaries:voice}
The process to extract audio from each video involve five steps.
(1)~We download YouTube videos in MPEG-4 format. (2)~We extract the corresponding stereo audio channels using the ffmpeg tool at 44.1KHz sampling rate.
(3)~Both audio channels are then mixed and down-sampled to 8KHz, using the sox utility; aiming to match same acoustic conditions between the YouTube audio and the training samples employed to develop the following audio analysis modules. 
(4)~We rely on Voice Activity Detection (VAD) to discriminate non-speech audio segments for further processing. (5)~We use Automatic Speech Recognition (ASR) to perform the speech-to-text transcription. Note that previous systems were originally trained and tuned using conversational telephone speech.

\descr{Voice Activity Detection.} 
VAD is often used as an upstream processing step intended to prevent unwanted data from entering later stages.
The VAD system we use is based on~\cite{balcells2016} and uses long short-term memory (LSTM) recurrent neural networks. %
We train and evaluate it using call center audio, 20 hours and 1.4 hours respectively, with error rates ranging from 5\% to 8\%.%

\begin{figure*}[t] 
\centering
\subfloat[Non-raided.]{\includegraphics[width=.48\columnwidth]{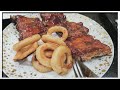}\label{fig:thumbnails:A}}
~
\subfloat[Non-raided.]{\includegraphics[width=.48\columnwidth]{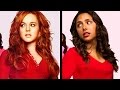}\label{fig:thumbnails:B}}
~
\subfloat[Raided.]{\includegraphics[width=.48\columnwidth]{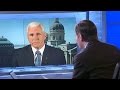}\label{fig:thumbnails:C}}
~
\subfloat[Raided.]{\includegraphics[width=.48\columnwidth]{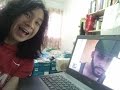}\label{fig:thumbnails:D}}
\vspace{-0.2cm}
\caption{Sample of thumbnails from our dataset.}\label{fig:thumbnails}
\end{figure*}

\descr{Automatic Speech Recognition.}
We use an ASR system for English from~\cite{Luque2017}, trained using the Kaldi toolkit~\cite{kaldi2008} and the Switchboard corpus~\cite{Godfrey1992}, which includes around 300 hours of conversational speech.
In particular, we adapt the Switchboard training recipe for nnet2 models from~\cite{kaldi2008}, and train two different systems.
The first one makes use of GMM/HMM models and speaker adaptation techniques. It employs triphone units with a total of 5,500 states and with a final complexity of 90,000 Gaussian mixtures. The second trains a vanilla DNN, composed of 4 hidden layers with 1,024 neurons each, on top of the alignments produced from the previous GMM system.

For the language modeling, we estimate a trigram language model using MIT Language Model Toolkit with Kneser-Ney Smoothing~\cite{Hsu08iterativelanguage}.
No lattice re-scoring is performed. 
The pronunciation dictionary, an orthographic/phonetic mapping, is from CMUdict, an open source pronunciation dictionary.\footnote{\url{http://www.speech.cs.cmu.edu/cgi-bin/cmudict}}
The target lexicon accounts for more than 40K words.
Note that neither ``bad words'' nor slang terms are in the original Switchboard lexicon.
To evaluate the ASR performance, we use a separated subset of the same Switchboard database accounting for 5 hours of speech.
The development results by the DNN based system, trained using only the Switchboard dataset, show a 13.05\% Word Error Rate (WER).
We finally run the DNN system on the Youtube audio dataset 
to generate the 1-best decoding transcription.
\descr{Evidence of controversial topics.}
Similar to what we did with the metadata, we also analyze the transcribed words to compare the different datasets.
We observe that most YouTube videos have a lot of verbal communication.
Specifically, 86\% of the videos have at least 10 words spoken with the median and average video transcription containing 317 and 1,200 words respectively. 
We also look at whether or not some terms  are more prevalent in raided YouTube videos, by averaging the \emph{TF-IDF} vectors separately for the two classes (raided and non-raided videos), and examining the most influential terms.
We find words like 
``black,'' ``police,'' ``white,'' ``shot,'' ``world,'' ``gun,'' ``war,'' ``American,'' ``government,'' and ``law'' in the top 20 terms in raided videos (in addition to some stop words and extremely popular terms that were excluded).
Of these, the only word that appears among the top 20 in the non-raided videos is ``government.''
The top terms for non-raided videos are different: they include words like ``god,'' ``fun,'' ``movie,'' and ``love.''

\begin{figure*}[t]
\centering
\includegraphics[width=0.92\textwidth, trim= 0 200 0 0, clip]{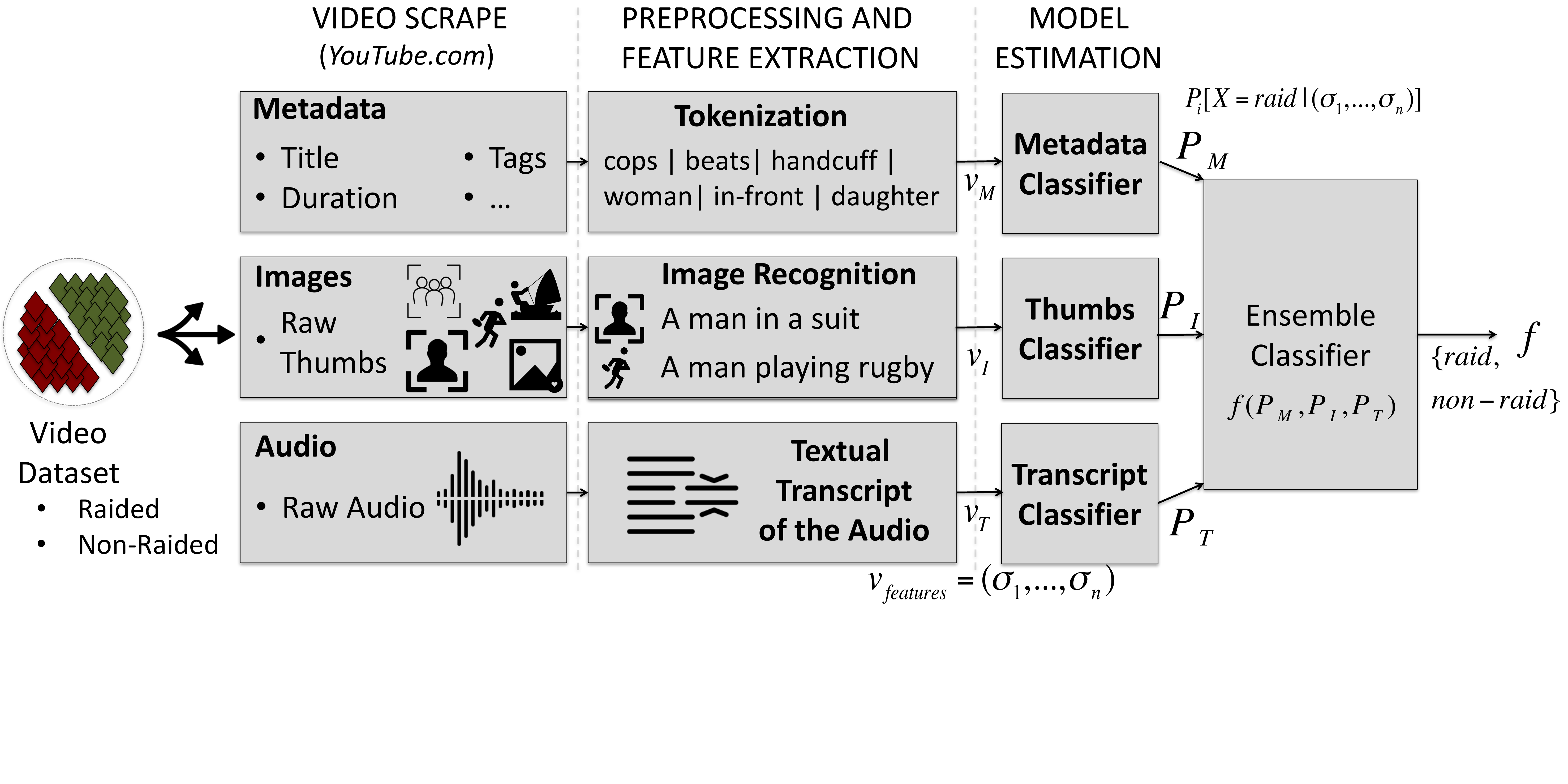}
 \vspace{-0.2cm}
\caption{Architecture of our proactive detection system.}\label{fig:architecture-system}

\end{figure*}

\subsection{Thumbnails}\label{preliminaries:thubms}

On YouTube, each video is also represented by an image thumbnail, used, e.g., in search results. 
Thumbnails provide viewers with a quick snapshot of the content of each video.
Although users can manually select them, %
by default, thumbnails are automatically selected from the video
and the user can choose one of three suggested options.

Using the YouTube API, we extract all available thumbnails from the videos in our dataset---specifically, using a combination of image recognition and content extraction tools (see below).
Note that %
thumbnails are not always available to download. %
In a few cases, this happens because the videos is not accessible via the API when gathering the thumbnails (which was done separately from the video itself, comments, and metadata), but, mostly, because the thumbnail was not been properly uploaded and is inaccessible even though the video is still available.

\descr{Image recognition.} To extract meaningful information from the thumbnails, we use deep neural networks~\cite{karpathy2015deep,vinyals2016show}.
A large corpus of images can be used to train a deep neural network: each image is annotated with visible context that is used as ground truth, and the resulting network can then recognize objects appearing in the images and generate an accurate description of them.

\longVer{We build on the work by~\cite{vinyals2016show} to train a generative model on top of a convolutional Neural Network (NN) and a language-generating recurrent NN.
The model is built from a very extensive dataset of annotated images (over 300,000) collected by Microsoft's COCO (Common Objects in Context) project.\footnote{\url{http://mscoco.org/}}
The system outputs a detailed description of the image provided as input. }

\descr{Context extraction.} For each thumbnail, we then output a description that represents the semantics involved in the image.
Figure~\ref{fig:thumbnails} shows images from four examples of different thumbnails, two in the \emph{raided} category and two in the \emph{non-raided} category.
The following descriptions have been automatically inferred from each of the images: 
(a) a white plate topped with a pile of food,
(b) a couple of women standing next to each other,
(c) a man in a suit and tie standing in front of a TV, and
(d) a woman sitting in front of a laptop computer. 
Note that each caption extracted not
only identifies the main actor within the picture (a plate, a couple of women, or a man), 
but also the background activity. 
However, these descriptions are automatically inferred based on a model bounded by the objects in the training images-set, thus, there might be misinterpretations.
\descr{Evidence of controversial topics.}
We use \emph{topic-modeling} to abstract the descriptions obtained from the images using ConceptNet~\cite{speer2017conceptnet}. 
In particular, we extract categories related to \emph{sports}, \emph{joy}, \emph{underage}, \emph{gender}, etc.
For example, some of the words in the \emph{joy} category are ``happy,'' ``smile,'' ``wedding,'' or ``Christmas.'' 
Table~\ref{tab:typeImages} shows a summary of the prevalence of words related to any of these categories across videos in our dataset.
We observe that there are a number of common topics displayed across both classes (e.g., \emph{nature}).
Likewise, female gender references are almost equal in both classes, with a slight bias towards \emph{raid} videos. 
Interestingly, the case of male gender references is clearly biased towards \emph{raided} videos, with males appearing in about 52\% of the non-raided videos and in 59\% of the raided ones.
Reference to clothes (e.g., ``tie'', ``dress'', ``jacket'', ``uniform'') is the most distinctive category with a 7.9\% difference between each class.

\begin{table}[t]
\centering
\vspace{0.2cm}
\small
\begin{tabular}{lrrr}
  \toprule
 {\bf Type}       &  {\bf Non-Raided} & {\bf Raided} & {\bf Diff.} \\ 
\midrule
  Clothing      &  25.5\%  &  33.4\% &   7.9\%\\
  Male-Gender&  52.4\%  &  59.1\% &  6.7\%\\
  Device     &  44.3\%  &  50.7\% &  6.4\% \\
  Vehicle      &   8.9\%  &  12.4\% &3.4\%  \\
  Animal     &  9.2\%  &  5.8\% &    3.4\% \\
  Sport      &  22.6\%  &  20.3\% &  2.2\% \\
  Color      &  12.5\%  &  10.7\% &  1.8\% \\
  Joy      &  1.6\%  &  2.8\% &      1.2\% \\
  Culture      &  1.6\%  &  0.7\% &  0.9\% \\
  Food      &  2.4\%  &  1.6\% &    0.8\% \\
  Female-Gender &  9.8\%  &  10.3\% &0.5\% \\
  Nature &  6.8\%  &  6.8\% & 0.1\% \\
\bottomrule
\end{tabular}
\vspace{-0.2cm}
\caption{Topics found across videos with thumbnails.}
\label{tab:typeImages}
\end{table}

This indicates that there are a number of thumbnails whose context can be used to characterize videos that could 
potentially be raided. However, numbers also indicate that 
thumbnails alone might not be able to fully model the differences 
between the two classes.
Our intuition at this point is that thumbnails can contribute towards the classification decisions, but they will not outperform other feature sets.
\review{While an exploratory analysis such as the one presented in this section can provide a general impression of the choice of models, it is hard to know upfront which model will better capture features singling out hate attacks. 
Extensive research work has covered the benefits of combining together different models~\cite{sewell2008ensemble}. 
Having different feature sets is generally the main reason for using a combination of classifiers. 
This is because different classification methods might perform better with a specific sub-set of features~\cite{jain2000statistical}. 
However, as pointed out in~\cite{sewell2008ensemble}, there is no definitive taxonomy of combined learning;  thus, an empirical comparison is paramount to determine the relative benefits and drawbacks in our domain.}

\section{Proactive Detection}\label{sec:methodology}

We now introduce our approach to provide a \emph{proactive} 
detection tool for videos targeted by hate attacks on online 
services, and on YouTube in particular. %
Our goal is to systematize this task, %
relying on a set of machine learning classifiers, each of which focuses on a different set 
of features extracted from online videos. 
Overall, we detail the set of features we use,
motivated by the findings reported in the previous section. 
\subsection{Overview}\label{sec:methodology:overview}

A high-level description of our detection system is presented in  
Figure~\ref{fig:architecture-system}. The system is first trained 
using the dataset videos as explained earlier. %
\smallskip

\noindent (1)~A set of prediction models $\mathcal{C}$$=$$\set{C_1, \ldots, C_i}$ 
that output a probability 
$C_i (\sigma_1, \ldots, \sigma_n)$$=$$P_i[Y$$=$$\texttt{raid}$$\mid$$(\sigma_1, \ldots, \sigma_n)]$
of each video $Y$ being raided given a feature vector $(\sigma_1, \ldots, \sigma_n)$ 
obtained from different elements $i$ of the video.
These models are referred to as \emph{individual classifiers}.\smallskip

\noindent (2)~A weighted model 
$f(\mathcal{C}) = \sum w_i \cdot C_i $
that combines all predictions in $\mathcal{C}$, where each of the 
classifiers $C_i$ is weighted by $w_i$ based on the performance obtained 
on a validation set. This set is different from the training set 
(used to build the individual probabilities) and the testing set (used to measure the efficacy of the classifiers). 
The process of weighting the individual predictors also 
serves as a way to calibrate the output of the probabilities. 
The final classifier will then output a decision 
based on its voting algorithm. 

To ease presentation, we refer to the model presented in (2)~as 
\emph{weighted-vote}. One can simplify the model by giving equal weight 
to all $w_i$ (typically $w_i = 1$) and obtaining a nominal value for $C_i$
before voting. In other words, applying a threshold for each $C_i$ (e.g.,
$0.5$) and creating an equal vote among participants. We refer to this 
non-weighted voting system as \emph{majority-vote}. One can further 
simplify the scheme by combining each individual prediction using the 
arithmetic mean of the output the probabilities---this is known as an 
\emph{average-prediction} system. 

Note that the parameters (i.e., $w_i$, $\epsilon$, and the thresholds 
for deciding the class in each $C_i$) used in both \emph{majority-vote} 
and \emph{average-prediction} are fixed and do not require calibration. 
Thus, the validation set is not used in these two modes.

\subsection{Feature Engineering}\label{sec:methodology:fengineering}

In the following, we discuss how we create the features vectors used by the different classifiers.
Our system extracts features from three different sources: 
(1)~structured attributes of the \emph{metadata} of the video, 
(2)~features extracted from raw \emph{audio}, and 
(3)~features extracted from raw \emph{images} (thumbnails).  
Based on the preprocessing described earlier, %
we transform non-textual elements of a video (i.e., audio and images) 
into a text representation. Other textual elements such as the title of 
the video and the tags are kept as text. 
These textual representations are then transformed into a fixed-size 
input space vector of categorical features. This is done by tokenizing 
the input text to obtain a nominal discrete representation of the words 
described on it. Thus, feature vectors will have a limited number of 
possible values given by the bag of words representing the corpus in the 
training set. When extracting features from the text, we count the frequency with which a word appears in the text. 

Since in large text corpus certain words--e.g., articles--can appear 
rather frequently without carrying meaningful information, we 
transform occurrences into a score based on two relative frequencies
known as \emph{term-frequency} and \emph{inverse document-frequency} 
(\emph{TF-IDF}s). Intuitively, the term frequency represents 
how ``popular'' a word is in a text (in the feature vector), and the 
inverse document-frequency represents how ``popular'' a word appears, provided that it does not appear very frequently in the corpus (the feature space). More formally, we compute as 
$idf(t) = log\frac{1 + n_s}{1 + df(s, t)} + 1,$
\noindent where $n_s$ is the total number of samples and $df(s,t)$ 
is the number of samples containing term $t$.

As for the thumbnails, after extracting the most representative descriptions per 
image, we remove the least informative elements and only retain 
entities (nouns), actions (verbs), and modifiers (adverbs and adjectives). 
Each element in the caption is processed to a common base to reduce 
inflectional forms and derived forms (known as stemming). 
Further, we abstract the descriptions obtained from the images using 
\emph{topic-modeling} as described earlier.

In our implementation, we extract features %
only from the thumbnail of a video. Again, this is mainly 
because the thumbnails are purposely selected %
to and encapsulate semantically relevant context.
However, we emphasize that our architecture could support the extraction of features from 
\emph{every} frame in the video.

\subsection{Prediction Models}\label{sec:methodology:models}

We use three independent classifiers to estimate the likelihood of a video being targeted by hate attacks.
These are built to operate independently, possibly when a new video is uploaded.
In particular, each classifier is designed to model traits from different aspects of the video. %
Available decisions are later combined to provide one unified output. %

We use three different classifiers, in an ensemble, because features obtained from different parts of a video are inherently 
incomplete, as some fields are optional and others might not be meaningful for certain video.
For instance, a music video might not report a lengthy 
transcript, or a thumbnail might not contain distinguishable context. 
Since any reliable decision system should be able to deal with 
incomplete data, ensemble methods are well-suited to this setting. 
Moreover, ensembles often perform better than
single classifiers overall~\cite{Dietterich2000EnsembleMethods}. 
The goal of each classifier is to extrapolate those controversial characteristics that, if present, are making the video likely to be raided.

\subsubsection{Metadata and thumbnail classifiers}
\label{sec:methodology:meta}

We build a prediction model such that $P_i(Y = \emph{raid})$ based on 
the features extracted from the metadata ($P_M$) and from the image thumbnails ($P_I$).
The architecture of these two predictors is flexible and accepts a range of classifiers. 
Our current implementation supports Random Forests (RFs), Extra Randomized Trees (XTREEs), 
and Support Vector Machines (SVM), both radial and linear. 
We opt for RF as the base classifier for  
$P_T$ and SVM with linear kernel for $P_M$. 
Both SVM and RF have been successfully applied to different aspects of 
security in the past (e.g., fraud detection~\cite{bhattacharyya2011data}) 
and have been shown to outperform other classifiers (when compared to 
180 classifiers used for real-world problems~\cite{Fernandez-Delgado:2014:WNH}).

\subsubsection{Audio-transcript classifier}\label{sec:methodology:voice}
Before feeding the transcripts to the classifier, we remove words that have a transcription confidence $p_{trans}$$<$$0.5$, as they are likely incorrectly transcribed (including them only confuses the classifier).
Note that this only removes 9.2\% of transcribed words.
Additionally, we remove  repeated terms that are mostly exclamations such as ``uh uh''  or ``hm hm''.
Finally, the transcripts contain tags for non-verbal communication such as noise, laughter, etc., 
which we leave in the text as they do carry predictive power.
\longVer{

\begin{figure}[t]
\centering
\includegraphics[width=.9\columnwidth, trim=0 350 250 0, clip]{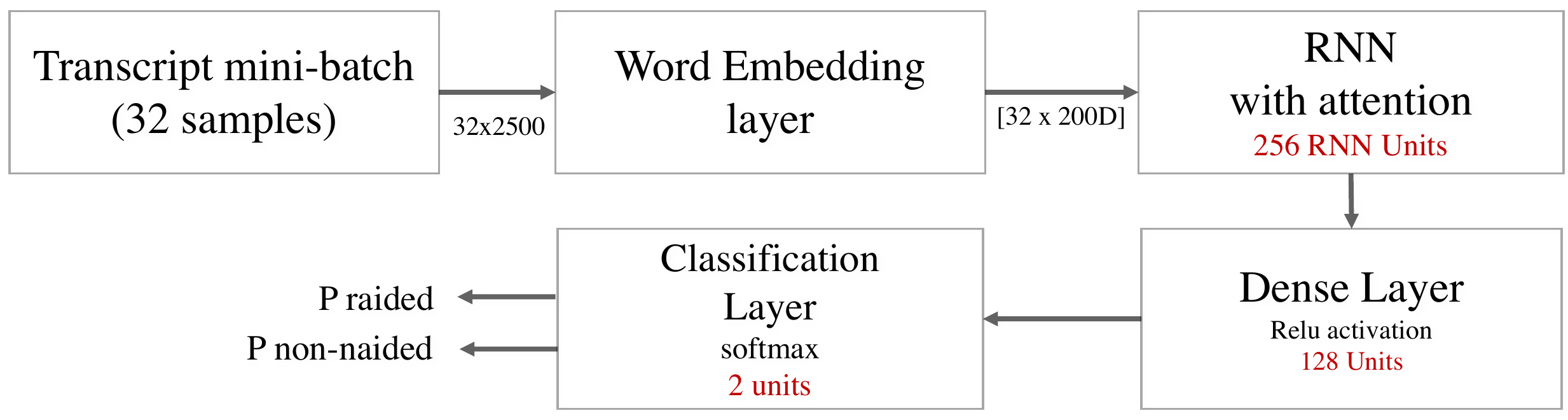}
\caption{Architecture of the transcript classifier. An Recurrent neural network with attention and pre-trained word embeddings was used.}\label{fig:deep-learning}
\end{figure}

}
We tried classifying using traditional TF-IDF based approaches, Convolutional Networks, and Recurrent Neural Networks (RNN), ultimately opting for 
the latter since it yields the best performance and is quite effective at understanding sequences of words and interpreting the overall context.
We also use an attention mechanism~\cite{bahdanau2014neural} that helps RNNs ``focus'' on word sequences that might indicate potentially raided videos. 
\longVer{An illustration of the overall architecture is shown in Figure~\ref{fig:deep-learning}.}
We also use GloVe to embed words into 200-dimensional vectors as we have relatively few training examples. 

\longVer{After empirical evaluation, we use a layer of 256 GRU units (with attention) and a recurrent dropout with $p{=}0.5$ for the RNN, which is connected to a fully-connected dense layer of 128 units with ReLu activation, and a softmax activation used to perform the binary classification.}

\subsubsection{Ensemble classifier}\label{sec:methodology:ensemble}
The objective of the ensemble method is to aggregate the predictions 
of the different base estimators. Each classifier individually models 
the likelihood that a video will be be targeted by hate attacks  
based on its set of features. %
The ensemble combines these decisions to make a more informed prediction. This %
allows for more robust predictions (in terms of confidence) %
and can result in a more accurate prediction. 
Our classifier is a stacking ensemble architecture, known to perform better than individual classifiers~\cite{wolpert1992stacked}. This architecture has been successfully used in other classification fields, for instance, credit scoring~\cite{wang2011comparative} and sentiment analysis~\cite{al2017using}.
We design our ensemble method to take a weighted vote of the available predictions.
To compute the best-performing set of weights, we estimate a function $f$ that takes as input each of the individual probabilities and outputs the aggregated prediction.
During training this function learns from an independent 
testing set, and will be used during testing to weight each 
prediction model $P_i$. Formally,  
$f(P_M, P_I, P_T) = \{\texttt{raid}, \texttt{non-raid}\}.$

For the decision function $f$ of our \emph{weighted-vote} 
algorithm (see the overview paragraphs at the beginning of this section), %
we use a distribution function 
that models how an expected probability in the testing set is 
affected by individual decisions $P_i$ in a multiple regression. 

In our implementations, we use different underlying 
classification algorithms for estimating $f$. 
However, in the next section, we only present the results for each of the individual classifiers and 
two ensemble methods, namely,
\emph{weighted-vote}, and 
\emph{average-prediction}. For the former, weights are 
fit using XTREE~\cite{geurts2006extremely}. %
For the latter, we also test different settings. %
In particular, we try settings where one of the classifiers is given a fixed weight of $w = 0$ and we average the others. 

\begin{table*}[t]
\centering
\setlength{\tabcolsep}{4pt}
\small
\begin{tabular}{llccc}
\toprule
{\bf ID} & {\bf Description} & {\bf Training} & {\bf Validation}  & {\bf Test}\\
\midrule
\experimentA & Random YouTube vs. all on 4chan & 731+731 & 243+243 & 13,470+244 \\[0.5ex]
\experimentE & All non-raided vs. raided on 4chan & 258+258 & 85+85 & 14,890+85 \\ [0.5ex]%
\experimentD & Non-raided on 4chan vs.  raided on 4chan& 258+258 & 85+85 & 446+85 \\
\bottomrule
\end{tabular}
\vspace{-0.2cm}
\caption{Number of samples used in our experiments. The sets are balanced as there is the same amount of samples per each class (\textit{class 1 samples+class 2 samples}) in training and validation, while they are unbalanced in the test set.} 
\label{tab:samples}
\end{table*}

\section{Evaluation}\label{sec:evaluation}

In this section, we present the setup and the results of our experimental evaluation. %

\subsection{Experimental Setup}
\label{sec:evaluation:setup}

We aim to show that we can distinguish between raided and non-raided videos.
However, there are several subtasks we also want to evaluate, aiming to better characterize the problem and understand how our classifiers perform. 

\descr{Experiments.} We start by trying to distinguish between random YouTube videos and those that are linked from \dspol. Next, we distinguish between the videos that are raided and those that are not (whether posted on \dspol or not). Finally, we predict whether videos posted on 4chan will be raided.

More specifically, in \experimentA, we test if our classifiers are able to distinguish between videos linked from \dspol and a random video uploaded to YouTube, aiming to gather insight into the ability to discriminate between videos {\em potentially} raided vs. those that are not at risk at all. 
Then, \experimentE evaluates whether or not the classifier can distinguish between any non-raided video (i.e., regardless of whether it is a random YouTube video or one posted on 4chan) and videos that will be raided.
Finally, in \experimentD, we focus on videos posted on 4chan, and determine which are going to be raided and which are not;
this ensures that we can not only predict whether a video was posted on 4chan, but whether or not the video will be raided.

\descr{Train, Test, and Validate Splits.}
We split our datasets into three chunks: two for training and tuning parameters of the ensemble (training and validation) and one for testing,
and report performance metrics on the latter.
As we are dealing with highly unbalanced classes (there are multiple orders of magnitude more videos posted to YouTube than those posted to 4chan, let alone those that are raided), we balance the training and validation sets to model both classes properly, but leave the test set unbalanced.
We have tried to train and tune the classifiers using unbalanced sets, but preliminary results have shown that this setup does not allow an accurate classification. This setup underperforms \review{by at least 0.12 AUC in all the different experiments} compared to the results we had by training and tuning on balanced sets.
The test set, however, remains unbalanced to more realistically model the ratio of raided videos against non raided videos in the wild.

The total number of videos in each split is proportionally sampled depending on the less populated class, assigning splits of 60\%, 20\%, and 20\% to the training, validation, and test sets. The more populated class uses the same amount of samples for training and validation, while it will have all the remaining samples in the test set. This procedure is repeated ten times and the results are averaged over the ten different rounds.
We decided to use random sampling for the training set, rather than a stratified sampling based on the categories of the videos as we believe this allows us to present results from a worst-case scenario: training videos may not fully represent test videos and, as consequence, our classifiers may perform slightly worse.
Table~\ref{tab:samples} summarizes all settings in our experiments, along with the number of samples used.

\descr{Evaluation Metrics.}
We evaluate our system using precision, recall, and F1-measure.
Overall, these metrics are a good summary of the performance of an 
classifier in terms of 
True Negatives (TN),
False Negatives (FN), 
False Positives (FP), and 
True Positives (TP); however, they are not ideal for comparing 
results across different experiments.
Therefore, we also plot the Area Under Curve (AUC), which reports the TP-rate ({\em recall}) against the FP-rate ({\em 1 - recall}).  

%

\begin{table*}[t]
\setlength{\tabcolsep}{0.2cm}
\centering
\small
\begin{tabular}{|l|c|c|c|c||c|c|c|c||c|c|c|c|}
\hline 
& \multicolumn{4}{c|}{\experimentA} & \multicolumn{4}{c|}{\experimentE} & \multicolumn{4}{c|}{\experimentD} \\
\hline
{\bf Classifier} & {\bf PRE}  & {\bf REC} & {\bf F1} & {\bf AUC} & {\bf PRE}  & {\bf REC} & {\bf F1} & {\bf AUC} & {\bf PRE}  & {\bf REC} & {\bf F1} & {\bf AUC}\\
\hline

transcripts & 0.05 & 0.60 & 0.10 & 0.79 & 0.03 & 0.56 & 0.06 & 0.79 & 0.32 & 0.58 & 0.40 & 0.73\\
metadata & 0.13 & 0.89 & 0.23 & \textbf{0.96} & 0.03 & 0.85 & 0.06 & \textbf{0.94} & 0.32 & \textbf{0.71} & 0.44 & 0.79\\
thumbnails & 0.02 & 0.64 & 0.05 & 0.62 & 0.01 & 0.66 & 0.02 & 0.61 & 0.18 & 0.55 & 0.27 & 0.56\\
\hline
\hline
weighted-vote ensemble & 0.12 & \textbf{0.91} & 0.21 & \textbf{0.96} & 0.03 & \textbf{0.88} & 0.05 & \textbf{0.94} & 0.34 & 0.69 & 0.45 & \textbf{0.80}\\
\hline
average-prediction ensemble & \textbf{0.15} & 0.85 & \textbf{0.26} & \textbf{0.96} & \textbf{0.04} & 0.82 & \textbf{0.07} & \textbf{0.94} & \textbf{0.35} & 0.69 & \textbf{0.46} & \textbf{0.80}\\
\hline
\end{tabular}
\vspace{-0.2cm}
\caption{Results for \experimentA, \experimentE, and \experimentD. PRE stands for precision, and REC for recall. The ensemble classifiers have different inputs: the weighted-vote classifier receives inputs from all three the individual ones, while the average-prediction does not receive the thumbnail classifier input.%
} 
\label{tab:exp1}
\label{tab:exp2}
\label{tab:exp3}
\label{tab:nothumb}
\vspace{-0.2cm}
\end{table*}

\subsection{Experimental Results}
\label{sec:evaluation:results}

We now report the results of our experimental evaluations, as per the settings introduced above. 
To ease presentation, we only report metrics for the individual classifiers as well as two ensemble methods: \emph{weighted-vote} and 
\emph{average-prediction}. 
We do not report results for other ensemble classifiers (\emph{simple-voting} and the other underlying algorithms for estimating the weights), since they under-perform in our experiments. 
For \emph{weighted-vote}, weights are fit using XTREE~\cite{geurts2006extremely}, 
as described earlier. 
Also note that, for \emph{average-prediction}, we find that the thumbnails classifier tends to disagree 
with the metadata and the transcripts classifiers combined. %
Therefore, for \emph{average-prediction}, we fix a weight of $w = 0$ 
for the thumbnails classifier (i.e., $w_{thumbnail} = 0$).

\descr{Experiment 1.} %
In this experiment we study whether we can predict that a video 
is linked from 4chan. 
Results are reported in Table~\ref{tab:exp1}. %
Since we are dealing with a rather unbalanced validation set (in favor of the negative class), it is not surprising that precision drops to values close to 0 even though we maintain high recall.%

Looking at the results obtained by the individual classifiers, 
\review{the weighted-vote ensemble classifier matched the best recall from the metadata individual classifier} (0.91). \review{The model relied almost entirely on the metadata classifier (weight equal to 0.987) while assigning very low weights to transcripts (0.007) and thumbnails (0.006).}
The best AUC is the same between the metadata classifier and the two ensemble classifiers (0.96). 

\begin{figure*}[t]%
\centering
\hspace*{-0.4cm}
\subfloat[Experiment 1]{
\includegraphics[width=.38\textwidth, trim=0 5 0 10, clip]{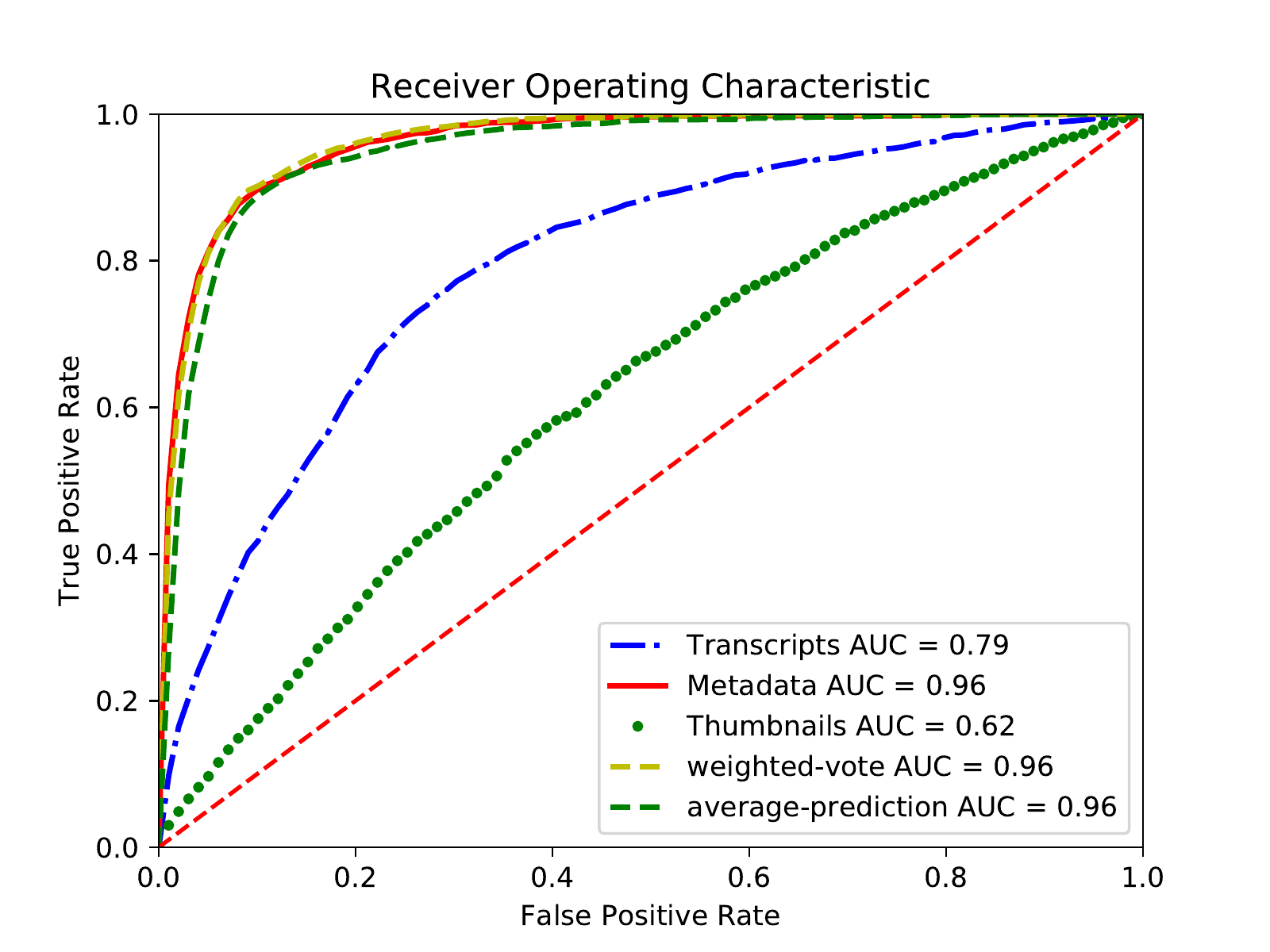}\label{fig:ROCA}
}
\hspace*{-0.75cm}
\subfloat[Experiment 2]{
\includegraphics[width=.38\textwidth, trim=0 5 0 10, clip]{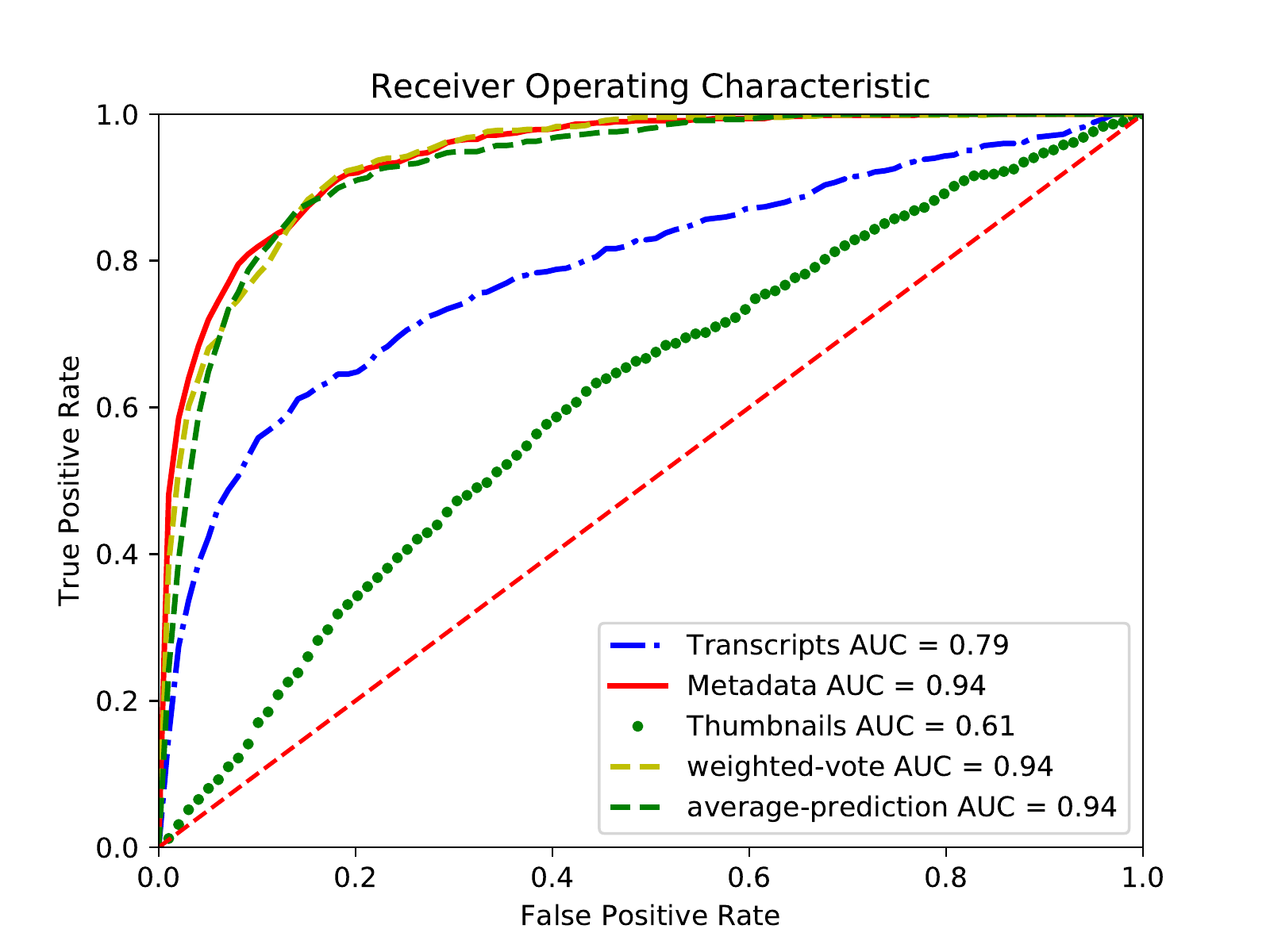}\label{fig:ROCE}
}
\hspace*{-0.75cm}
\subfloat[Experiment 3]{
\includegraphics[width=.38\textwidth, trim=0 5 0 10, clip]{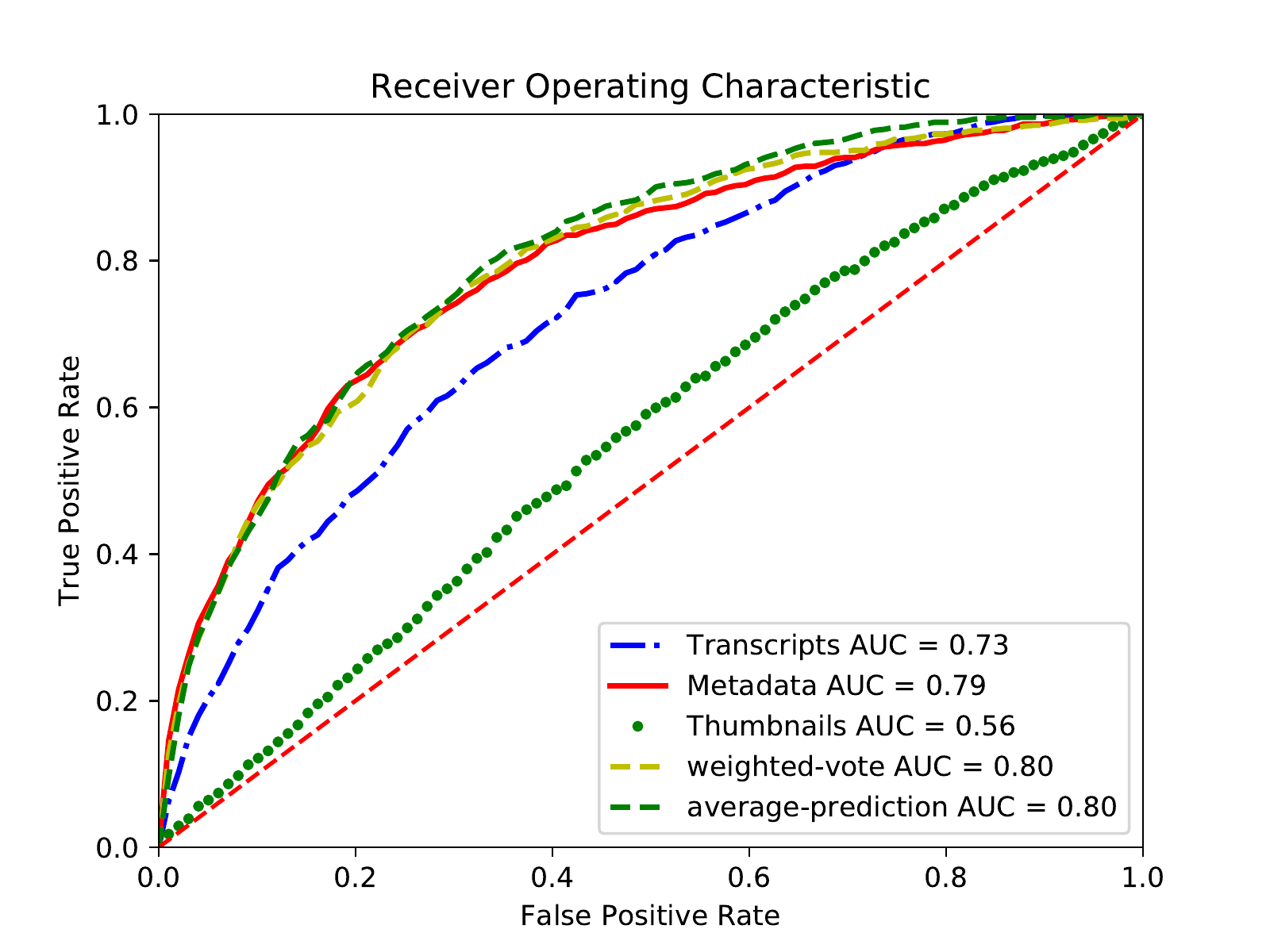}\label{fig:ROCD}
}
\vspace{-0.2cm}
\caption{ROC curves for each experiment. AUC values for Thumbnails, Transcripts, Metadata, and Ensemble classifiers, XTREE, and Average probabilities. \review{This metric provides a fair comparison across experiments.} }
\end{figure*}

In Figure~\ref{fig:ROCA}, we also plot the ROC curve for all five classifiers.
The individual AUC scores are 0.79, 0.96, 0.62 for the transcripts, metadata, and thumbnails, respectively, while the two ensembles ({\em weighted-vote} and {\em average-prediction}) score 0.96.
The {\em weighted-vote} ensemble has the highest AUC throughout most of the x-axis, although, the ROC curve essentially overlaps with that of the metadata classifier. 
The two ensembles have different strengths: the {\em weighted-vote} has the highest recall and AUC values, but the {\em average-prediction} (with $w_{thumbnail} = 0$) has highest precision, and F1-measure.

\descr{Experiment 2.} 
In Figure~\ref{fig:ROCE}, we report the AUC when classifying raided and 
non-raided videos---regardless of whether the latter are 
random YouTube videos or non-raided ones posted on 4chan. 
Unlike \experimentA, among the individual classifiers, the best performance is achieved by the audio-transcript classifier, except for recall. %
This setting also yields high recall (0.88) when 
combining all classifiers into the {\em weighted-vote} ensemble. As in \experimentA, the {\em weighted-vote} ensemble presents the highest recall and AUC, %
but the {\em average-prediction} has higher precision, and F1-measure. \review{Once again, the model relied almost entirely on the metadata classifier (weight equal to 0.984) while assigning very low weights to transcripts (0.008) and thumbnails (0.008).}
Figure~\ref{fig:ROCE} shows a similar situation as in the previous experiment: the ROC curve for the metadata classifier is really close to or overlapping with the ones for the two ensemble. AUC equals to 0.61 for thumbnails, 0.79 for transcripts, and 0.94 for metadata. Whereas, the two ensemble classifiers achieve 0.94 AUC as the metadata individual classifier

\descr{Experiment 3.} %
Finally, we evaluate how well our models discriminate between 
raided videos posted to 4chan and non-raided videos also 
posted to 4chan. 
Our results confirm that this is indeed the most challenging task. 
Intuitively, these videos are much more similar to each other 
than those found randomly on YouTube as \dspol is interested in a particular type of content.

This setting shows a clear case for the ensemble classification yielding appreciably better performance.
Overall, the individual classifiers, i.e., transcripts, metadata, and thumbnails, reach AUCs of 0.73, 0.79, and 0.56, respectively, whereas, both the ensemble classifiers reach 0.80.
Nevertheless, the ROC curve in Figure~\ref{fig:ROCD} 
shows how the {\em weighted-vote} ensemble is sometimes penalized by the weakest performing classifier (i.e., thumbnails classifier). 
This is apparent by comparing the {\em weighted-vote} and {\em average-prediction} (recall that $w_{thumbnails} = 0$ in the latter). \review{The weights assigned by the model confirm this: the metadata classifier's weight is 0.868, while the weight for transcripts is 0.034 and the one for thumbnails is 0.098. As expected, the metadata has a very important weight (although not as high as in the previous experiments), surprisingly, the thumbnails have a higher weight than the transcripts. This means that the transcripts classifier tend to agree with the metadata one in most of the cases they correctly flag the videos, while the thumbnails classifier identifies correctly some videos when the other classifiers fail.}

\subsection{Choosing an Ensemble} 
The goal of the our system is to make the best final ``decision'' possible given the choices made by the individual classifiers. In absolute terms, the \emph{weighted-vote} (with XTREE as baseline estimator) yields the best performance in all three experiments in terms of recall (and overall AUC). In particular, it outperforms the \emph{average-prediction} ensemble in two of the tasks: modeling videos from \dspol ({\sc Experiment 1}), and detecting raided videos in general ({\sc Experiment 2}). When restricting our attention  to the detection of raids of videos posted on 4chan ({\sc Experiment 3}), both ensemble methods are comparable in terms of recall. However, when looking at precision, we find that \emph{average-prediction} outperforms \emph{weighted-vote}. The trade-off between having good precision and good recall will have an important impact on the amount of vetting work required by providers deploying our system, as we discuss in more detail in Section~\ref{sec:discussion}. 

In the following, we provide an explanation as to why the two ensemble classifiers report similar results in some experiments and different ones in others. When using a base estimator to fit the best weights for the individual classifiers, we observe a bias towards the decisions made by the metadata classifier. This is expected, as this classifier is the one that performs best among the individual classifiers (and substantially so in both {\sc Experiment 1} and {\sc Experiment 2}). On the contrary, the thumbnails classifier performs worst, except for recall in {\sc Experiment 2}. 

Note that our data include videos with partially-available features.
When this happens, the ensemble classifier is forced to make a decision based on the other inputs. 
It is the thumbnails case, which are not always available. %
This is why we evaluated the \emph{average-prediction} ensemble method forcing a weight $w_{thumbnails} = 0$. 
In this setting, the \emph{weighted-vote} method with XTREE provided similar results, since XTREE initially assigned a low weight (although not exactly 0) to the thumbnails. 

Overall, with the \emph{average-prediction} method, precision, and F1-measure are always better than for %
the XTREE ensemble classifiers. %
This means that this configuration reduces the number of false positives and, as a consequence, is slightly more accurate. 
Therefore, when the individual classifiers have similar performance, the ensemble is better than the best options among the single classifiers. %

\section{Discussion}\label{sec:discussion}

Our experiments show that we can model YouTube videos and predict those likely to be raided by off-platform communities.
In other words, our results indicate that it is possible to develop automated techniques to mitigate, and possibly prevent, the socio-technical problem of online attacks and harassment.
What still needs to be ironed out is how our techniques could be integrated and deployed by online services like YouTube. %
Although devising a path to adoption is beyond the scope of this paper, we discuss this aspect later in this section.

\smallskip\noindent\review{\textbf{Data collection.} The data collection involved in this work presented a variety of challenges. %
For example, how can we be sure that the HCPS metric is effective in identifying hateful comments?
How do we know that non raided videos on \dspol or random videos are not False Negative cases?
How can we ensure that videos are actually random to a reasonable extent?
Balancing these questions while aiming to support large-scale, automated analysis thus played a large role in our design decisions.}

\review{The use of HCPS and time lag metrics from \cite{4chan} allowed us to evaluate the system on ground-truth videos that were already discovered by previous work.
Unfortunately, as expected, some of these videos were quite popular, with an extremely large number of comments making manual checking not viable.
Conversely, using automatically quantified metrics may result in assigning the wrong class to a video (e.g., HCPS may not identify hateful comments in a specific video).
To mitigate this issue, we took a conservative approach, constraining our dataset to videos that, e.g., had a minimum HCPS and a time lag of less than a day.
While this does leave questions about ``borderline'' cases, it also results in less ambiguity surrounding our ground truth. }
\review{Also note that the random videos dataset was collected by category, aiming to reproduce the same distribution of categories as our ground truth.
As YouTube does not provide detailed statistics on the overall distribution of videos, we opted to shape our negative class following the worst possible case: an exact copy of the \dspol distribution.
That is, our negative class comprises videos that (at least in terms of category) \dspol might find ``interesting.''
Further, collecting the negative dataset via API queries removes additional concerns about human decisions influencing its composition.
A limitation of this approach is that, by relying on \dspol videos and the HCPS data from previous work, we were not able to apply the algorithm to the random videos.
However, such limitation may only have negative effects on our results, meaning that our experiments yielded a lower bound on the system efficiency.
}

\descr{Evaluation.} Our evaluation shows that we can reliably distinguish videos that are likely to be raided from regular YouTube videos. %
This means that YouTube could run our system at \emph{upload time}, determining the likelihood that a video will be raided at some point in the future.
The platform could then adopt mitigation strategies for ``risky'' videos, e.g., by manually moderating comments. %
This is a practice already employed by YouTube~\cite{techcrunchYoutube}, however, at the moment, the efficacy of this moderation has been questioned %
 due to the sheer volume of content, in addition to YouTube's focus on removing {\em inappropriate} videos rather than protecting users against raids.

Using our system, we estimate that only 16\% of videos would require any action---an estimation based on \experimentE. While this might appear to be a very large number, it is still less than having to monitor all videos.
Moreover, additional automated tools could be deployed to check whether a raid is actually occurring before being passed along for human review.
Furthermore, YouTube has pledged to hire 10,000 new workers to monitor content about a year ago~\cite{10kmoderators}, but deploying our system could reduce the need for this human workforce, or at worst, allow them to focus on higher impact content.

In addition, \experimentD demonstrates that, when provided with videos linked from fringe communities such as \dspol, our system can successfully identify videos likely to be raided with reasonable accuracy.
This is a much more difficult task, and thus the correct detections are less than before, since videos are very often being posted to \dspol without the actual intent of having raiders show up in the first place.
Furthermore, the number of videos posted on \dspol is much smaller than those uploaded to YouTube overall---for instance, the \dspol dataset from \cite{4chan} contains links to 93k YouTube videos posted over 3 months.
Among these, we only selected those that had clear evidence of raids by restricting the thresholds of HCPS and time lag (see Section~\ref{sec:raids}), %
also discarding videos which could have been raided from the non-raided group.
This choice is extremely conservative (yielding 428 videos), aiming to have a reliable ground truth on which to evaluate the system.
Although we could have relaxed our thresholds and obtained better results overall, the applicability to real-world use cases would likely have been affected, as our ground truth dataset would have included videos that had controversial or combative statements, but were not actually raided. 

\descr{Adoption.} Our system is really geared to identify videos that are \emph{at risk} of being raided, thus, YouTube could use it as a {\em filter} --- flagging videos that are risky and/or might require particular attention. 
\review{We emphasize that an automated system such as the one presented in this paper inherits classical machine-learning limitations, e.g., dealing with misclassifications.
However, the threshold for what is considered actionable can be tuned, since in the end the classifiers output a probability as opposed to a purely binary classification.
In other words, a deployed system could be adjusted to focus on minimizing false negatives (i.e., videos that will be raided but are misclassified by the system) while still maintaining high recall.}
Tuning the model to balance the overall number of flagged videos as well as ensuring that they are indeed at high risk can help reduce the impact of false positives.
Given that our datasets are extremely unbalanced, high precision, on the other hand, is not a top priority.
As mentioned above, the system would flag videos likely to be raided, thus, helping to tackle the problem of aggression by reducing the videos that need to be monitored as high risk ones.
\review{Overall, we envision the system not as an end-all be-all solution, but as an early warning system, enhancing other mechanisms the platform decides to put in place
At minimum, e.g., allowing them to focus human efforts (content moderators etc.) their efforts where abusers are most likely to attack.}
\review{While this would certainly reduce the amount of human labor involved in dealing with raids, it could also be used to focus more expensive algorithmic systems on high risk videos.
E.g., by introducing a post-prediction system leveraging the HCPS metric discussed earlier (c.f. Section~\ref{sec:raids}) as well as other existing mechanisms to flag offending comments~\cite{chatzakou2017mean}. 
}

\smallskip\noindent\review{\textbf{Limitations.} As mentioned previously, our data collection attempted to minimize the mislabeling of videos linked on \dspol.
This approach allowed us to be reasonably confident that the raided videos dataset contained only videos attacked by the \dspol community.
However, these constraints limited the size of our dataset (out of the more than $5,000$ videos linked on \dspol we considered only $428$ videos as having been raided).
This relatively small dataset was a limiting factor in the performance of our classifiers.}

\review{Moreover, we assume that negative class dataset does not contain mislabeled samples.
The assumption may not hold up 100\% of the time since labeling errors occur with both manual and automatic checks.
That said, while mislabeling in our negative class dataset might affect classifier performance to some degree, the overall validity of our system is not affected.}

Also note that \dspol, though a very good example of a tight-knit community used to coordinate and disrupt other social groups, is not the only community responsible for performing raids against YouTube videos. 
Other Web communities, e.g,. Reddit%
~\cite{incels} or Kiwi Farms\footnote{\url{https://kiwifarms.net/}} %
 also regularly take part in raiding activity.
The same techniques presented here, however, can be used to detect raids from other communities.

\descr{Raids and communities.} Finally, it might be tempting to dismiss the relatively low occurrence of raids, vis-\`a-vis the number of YouTube videos posted every day, as being a niche problem. 
On the contrary, harassment and bullying on YouTube are widely recognized as a serious issue by authorities on the matter%
~\cite{nobully}, and news reports are filled with ghastly stories%
~\cite{jesseslaughter}
 and advice on how to deal with hateful and harassing YouTube comments in particular%
~\cite{makeupgirl,youtubestars}.

Although we are not aware of any suicide cases directly linked to YouTube raids, victims have indeed been active on YouTube %
\cite{gayrightssuicide} %
and thus raids pose very serious safety risks.
Overall, even if the majority of content on YouTube (and other social media platforms) tends to be ``safe,'' we should not discard the outsized effects that this negative behavior has.
From a purely pragmatic point of view, advertisers providing the primary income stream for sites like YouTube have been re-thinking their reliance on social media in light of the recent surge in anti-social behavior%
~\cite{youtubeadvertisingboycott}.
From a societal point of view, raiding behavior is a pressing concern; it is a direct threat to free speech and civil discourse, and causes emotional distress that can lead to dire consequences.
The efforts of the research community have enabled the long tail of the Web to succeed, building technologies that democratized information and shrunk the world.
Thus, while raids on YouTube videos do occur in the long tail, we argue that dismissing them as being too rare is an abdication of our social responsibility.

\section{Conclusion}\label{sec:conclusions}

This paper presented a supervised learning based approach to automatically determine whether a YouTube video is likely to be ``raided,''
i.e., receive a sudden spike in hateful comments as a result of an orchestrated effort coordinated from another platform.
Our experimental results showed that even single-input classifiers that use metadata, thumbnails, or audio transcripts can be effective,
and that an ensemble of classifiers can reach high detection performance, thus providing a deployable early-warning system. 

Overall, our work represents an important first step toward providing video platforms like YouTube with proactive systems geared to detect and mitigate coordinated hate attacks. We discussed potential deployment strategies that could be taken by YouTube (or other providers), i.e., running our tool on every video at upload time and/or monitoring fringe communities such as 4chan to screen videos that are linked to on those platforms. 

Note that the classifiers presented in this paper are not meant to provide a mechanism for {\em censoring} content or users, nor to identify users possibly involved in raids. Rather, we aim to identify content that is at risk of attack; once identified, \emph{proactive} solutions to protect against raiders can be taken by the service providers.
While the specifics are beyond the purpose of this paper, we believe that there are actions that can be taken that protect freedom of expression while also preserving civil discourse.
For example, temporarily disabling or rate limiting comments, requiring new comments to be approved before going live, or simply notifying the poster that a raid might be coming could serve to balance protection vs. expression.

As part of future work, we plan to use \review{rank aggregation techniques as ensemble, as well as} deep-learning methods to fuse audio, video, and metadata into a single classifier.
\review{This design follows a different approach with respect to the current one. It needs more data (and as consequence more time and computational resources for training) than the amount of raided videos we currently have, but, as it can manage the relations among the different features types, it is likely to have better performances once trained properly.
}
We also plan to look into raids from other communities, such as Reddit, Gab.ai, and Kiwi Farms. 

\descr{Acknowledgments.} This project has received funding from the European Union's Horizon 2020 Research and Innovation program under the Marie Sk\l{}odowska-Curie ENCASE project (GA No. 691025). Enrico Mariconti was also supported by the EPSRC under grant 1490017.

\small

\bibliographystyle{abbrv}
\end{document}